\documentclass{ws-procs9x6}

\def\bnabla{\mbox{\boldmath$\nabla$}}
\def\bPsi{\mbox{\boldmath$\Psi$}}
\def\bv{\mbox{\boldmath$v$}}
\def\bb{\mbox{\boldmath$B$}}
\def\be{\mbox{\boldmath$E$}}

\def\bOne{\mbox{\boldmath$1$}}
\def\bF{\mbox{\boldmath$F$}}
\def\bx{\mbox{\boldmath$x$}}
\def\dA{d\mbox{\boldmath$A$}}
\def\dl{d\mbox{\boldmath$l$}}

\def\la{\leftarrow}
\def\ra{\rightarrow}
\def\lra{\leftrightarrow}
\def\qra{{q\ra}}
\def\qla{{\la q}}
\def\rra{{r\ra}}
\def\rla{{\la r}}

\begin{document}


\title{The Long Term: Six-dimensional Core-collapse Supernova Models
}





\author{C.~Y.~CARDALL$^{1,2}$, A.~O.~RAZOUMOV$^{1,2,3}$, E.~ENDEVE$^{1,2,3}$, and A.~MEZZACAPPA$^1$ }

\address{$^1$Physics Division, \\ Oak Ridge National Laboratory, \\ 
Oak Ridge, TN 37831-6354, USA}
\address{$^2$Department of Physics and Astronomy, 
University of Tennessee, 
	Knoxville, TN 37996-1200, USA}%
\address{$^3$Joint Institute for Heavy Ion Research, \\
        Oak Ridge National Laboratory, \\
        Oak Ridge, TN 37831-6374, USA}

\maketitle

\abstracts{
The computational difficulty of six-dimensional neutrino radiation hydrodynamics has spawned a variety of approximations, provoking a long history of uncertainty in the core-collapse supernova explosion mechanism. 
Under the auspices of the Terascale Supernova Initiative, we are honoring the physical complexity of supernovae by meeting the computational challenge head-on, undertaking the development of a new adaptive mesh refinement code for self-gravitating, six-dimensional neutrino radiation magnetohydrodynamics. 
This code---called {\em GenASiS}, for {\em Gen}eral {\em A}strophysical {\em Si}mulation {\em S}ystem---is designed for modularity and extensibility of the physics. 
Presently in use or under development are capabilities for Newtonian self-gravity, Newtonian and special relativistic magnetohydrodynamics (with `realistic' equation of state), and special relativistic energy- and angle-dependent neutrino transport---including full treatment of the energy and angle dependence of scattering and pair interactions.
}

\section{The Challenges of Core-collapse Supernovae}
\label{sec:challenges}

In taking stock of `long-term' efforts to understand core-collapse supernovae, we reflect upon the fact that supernovae have been challenging us for centuries. Their very existence helped overturn worldviews.  Their explosion mechanisms and remnants involve all four fundamental forces, and many (if not most) branches of physics. 
A cornucopia of electromagnetic radiation observables continues to provide intriguing puzzles, and yields some clues regarding the violent proceedings of a massive star's death. 
But direct observational penetration of the secrets of neutron star birth and initiation of the explosive ejection of the stellar envelope demands extraordinary efforts aimed at the detection of gravitational waves and neutrinos, the only messengers carrying direct information from the extreme conditions of a newly-collapsed stellar core. 
And the associated theoretical penetration---required for both the prediction and interpretation of expected gravitational wave and neutrino signals---comprises
algorithmic and computational issues that will challenge computational physicists and tax state-of-the-art supercomputers for years to come.

In western civilization, supernovae played a role in changing prevailing notions of the universe in at least two eras. Remarkably, of the handful of supernovae in our Milky Way Galaxy recorded by humanity, two were observed by Tycho (1572) and Kepler (1604). Tycho's detailed observations established that the `new and never previously seen star' of 1572---and also another transient celestial phenomenon, a comet of 1577---were beyond the moon's orbit, `new phenomena in the ethereal world,' contributing to the overthrow of the Aristotelian worldview that included immutable heavens. In modern times, supernovae figured in the debate over whether 
the spiral nebulae 
were separate galaxies, each an `island universe' comparable to our Milky Way. It was recognized 
that the `novae' or `new stars' seen in these nebulae would have to be much more luminous than typical novae occuring in our galaxy. 
The phrases 
``giant novae,'' novae of ``impossibly great absolute magnitudes,'' 
``exceptional novae,'' and the German term 
``Hauptnovae'' or ``chief novae'' were used during the 
1920s.\cite{osterbrock01}
In a review article, Zwicky explained that it was deduced that `supernovae' 
were about a thousand times as luminous as `common novae,' and 
claimed that ``Baade and I first introduced the term `supernovae' in seminars and in a lecture course on astrophysics at the California Institute of Technology in 1931.''\cite{zwicky40}

Supernovae are classified by astronomers into two broad classes based on their optical spectra.\cite{filippenko97}
These classes are `Type I,' which have no hydrogen features, and 
`Type II,' which have obvious hydrogen features. These types have further 
subcategories, depending on the presence or absence of silicon and 
helium features in Type I, and the presence or absence of narrow 
hydrogen features in the case of Type II. In particular, 
supernovae of Type Ia exhibit strong silicon lines, 
those of Type Ib have helium lines, and those of Type Ic do not have 
either of these. Astronomers have also identified a number of distinct
characteristics in supernova light curves
(total luminosity as a function of time). 

There are two basic physical mechanisms for supernovae, but these do not line up cleanly with the observational categories of Type I and Type II. Type Ia supernovae are caused by a thermonuclear runaway that consumes an entire white dwarf, thought to be induced by accretion of matter from a companion star.
Supernovae of Type Ib, Ic, and II are produced by a totally different mechanism: the catastropic collapse of the core of a massive star. The observational distinctions of presence or absence of hydrogen or helium turn out to be unrelated to the mechanism; they depend on whether the outer hydrogen and helium layers of the star---which have nothing to do with the collapsing core---have been lost to winds or accretion onto a binary companion during stellar evolution. Of the two
physical mechanisms, core-collapse supernovae are the focus of the
present discussion.

We now consider the core-collapse supernova process in more detail.
Shortly after the discovery of the neutron in the early 1930s, Baade and Zwicky declared, 
``With all reserve we advance the view that supernovae represent the
transitions from ordinary stars to {\em neutron stars,} which in their final
stages consist of extremely closely packed neutrons.''\cite{baade34} 
This turned out to be true, at least for some `core-collapse' supernovae (those of Type Ib/Ic/II); a black hole is another possible outcome. 
The dominant fleshing-out of the core collapse process in the last two decades\footnote{See for example Ref. \cite{burrows90} for some information on earlier views of the mechanism.} 
has been
the delayed neutrino-driven explosion mechanism.\cite{wilson85,bethe85}

A core-collapse supernova results from the evolution of a massive star.
For most of their existence, stars burn hydrogen into helium. 
In stars at least eight times as massive as the Sun ($8\ M_\odot$), 
temperatures and densities become sufficiently high to burn 
through carbon to oxygen, neon, and magnesium; in stars of at least $\sim 10\ M_\odot$, burning continues through silicon to iron group elements. 
The iron group nuclei are the most tightly bound, and here burning in 
the core ceases. 

The iron core---supported by electron 
degeneracy pressure instead of gas thermal pressure, because of cooling
by neutrino emission from carbon burning onwards---eventually 
becomes unstable. Its inner portion undergoes homologous collapse
(velocity proportional to radius), and the outer portion collapses 
supersonically. 
Electron capture on nuclei is one instability leading to collapse, and 
this process continues throughout collapse, producing neutrinos.
These neutrinos escape freely
until densities 
in the collapsing core
become so high that even neutrinos are 
trapped. 

Collapse is halted soon after the matter exceeds nuclear density; 
at this point (``bounce''), a shock wave forms at the boundary between the homologous
and supersonically collapsing regions. The shock begins to move out,
but after the shock passes
some distance beyond the surface of the newly-born neutron star, 
it stalls as energy 
is lost to neutrino emission and endothermic dissociation of heavy 
nuclei falling through the shock.

It is natural to consider neutrino heating as a mechanism for
shock revival, because neutrinos dominate the energetics of
the post-bounce evolution.
Initially, the nascent neutron star is a hot thermal bath of dense nuclear matter, 
electron/positron pairs, photons, and neutrinos, containing most of 
the gravitational potential energy released during core collapse. 
Neutrinos, having the weakest interactions, are the most efficient 
means of cooling; they diffuse outward on a time scale of seconds, 
and eventually escape with about 99\% of the released gravitational energy.

Because neutrinos dominate the energetics of the system, 
a detailed understanding of their evolution will be integral to 
definitive accounts of the supernova process.
If we want to understand the origin of the explosion with energy $\sim 10^{51}$ erg, we cannot afford to lose (or gain) more than this amount during the period covered by the simulation. This requires careful accounting of the
neutrinos' much larger contribution to the system's energy budget. (For further discussion, and a review of work recognizing the importance of this point, see Ref. \cite{cardall05b}).

What sort of computation is needed to follow the neutrinos' evolution?
 Deep inside the newly-born neutron star, 
 the neutrinos and the fluid are tightly coupled (nearly in equilibrium);  but as 
 neutrinos are transported from inside the neutron star, they go from a nearly isotropic diffusive regime to strongly forward-peaked free-streaming. Heating 
 behind the shock occurs precisely in 
this transition region, and modeling this process accurately requires tracking both the energy and angle dependence of the neutrino distribution functions at every point
in space. 

A full treatment of this six-dimensional neutrino radiation hydrodynamics
problem is a major challenge, 
too costly for 
contemporary
computational resources. While much has been 
learned over the years through simulation of model systems of reduced dimensionality, there is as yet no robust confirmation of the delayed neutrino-driven scenario described above
 (see Sec. \ref{sec:history}).

Recent detections of a handful of unusually energetic Type Ib/c supernovae (often called `hypernovae') in connection with gamma-ray bursts
pose additional challenges to theory and observation. 
Prominent examples of this supernova/gamma-ray burst connection include SN1998bw / GRB980425,\cite{galama98,iwamoto98} SN2002lt / GRB021211,\cite{dellaValle03} SN2003dh / GRB030329,\cite{hjorth03,stanek03} and SN2003lw / GRB031203;\cite{thomsen04,cobb04,malesani04,galYam04} there are probably many others (see, for example, Ref. \cite{zeh04}). Like many gamma-ray bursts without direct evidence for a supernova connection, GRB030329 has evidence of a jet; GRB980425 and GRB031203 do not, and are also underluminous gamma-ray bursts (but still unusually energetic Type Ib/c supernovae).\cite{ghirlanda04,soderberg04} 
Determining the relative rates of jet-like hypernovae, non--jet-like hypernovae, and `normal' supernovae---and the possible associated variety of mechanisms---are 
important challenges.

In summary, the details of how the stalled shock is revived
sufficiently to continue plowing through the outer layers of the
progenitor star are unclear. In normal supernovae, it may well be that some combination of neutrino heating 
of material 
behind the shock, convection, and instability of the spherical
accretion shock leads to the explosion (see Sec. \ref{sec:history}). It is tempting to think that rotation 
(for example, Refs. \cite{fryer00,thompson04}) 
and magnetic fields 
(for example, Ref. \cite{wheeler04}) 
in more massive progenitors may play a more significant role in the rare jet-like hypernovae, perhaps giving birth to `magnetars,' the class of neutron stars with unusually large magnetic fields. This temptation appears to be sweetened by observational support.\cite{gaensler05,figer05} (Observations of two nearby supernova remnants may suggest that rotation and magnetic fields also operate in normal supernovae, perhaps subdominantly.\cite{burrows04b})

From the above discussion, several key aspects of physics that a core-collapse simulation must 
address can be identified; these are discussed in sections that follow, after a discussion of the history of approximate treatments of neutrino radiation transport and an overview of our new code.

\section{History of Neutrino Radiation Hydrodynamics}
\label{sec:history}

While in general terms supernovae have been challenging us for centuries, the challenge of their simulation via computer modeling has `only' been with us for a few decades---a `short term' in comparison with centuries, but still a `long term' in comparison with the time scales of individual academic careers.

Here we sketch the last two decades' progress on one critical aspect of core-collapse supernova simulations: the high dimensionality (three space and three momentum space dimensions---not to mention time dependence) of neutrino radiation hydrodynamics (see Table 1). The development of this aspect of the simulations is intertwined with important advances in the field, but of course does not represent every insight relevant to the explosion mechanism obtained via simulation or otherwise. 

\def\cal{}

\begin{sidewaystable}	
\tbl{Selected neutrino radiation hydrodynamics milestones in stellar collapse  simulations studying the long-term fate of the shock.
\label{history}}
{\begin{tabular}{@{}lcccccc}
\hline
Group & Year & Explosion & Total  & Fluid space  & $\nu$ space & $\nu$ momentum
	     \\
	     &       & & dimensions    & dimensions & dimensions & space dimensions \\
\hline
Lawrence Livermore\cite{bowers82,wilson85,bethe85} & 1982 & Yes$^*$ & 2 & 1 (PN)  & 1 & 1 (${\cal O}(v/c)$)    \\
\hline
Lawrence Livermore\cite{mayle85,wilson88}& 1985 & Yes$^*$ & ``2.25'' & ``1.5'' NS (PN)  & 1 & 1 (${\cal O}(v/c)$) \\
\hline
Florida Atlantic\cite{bruenn85,bruenn87,bruenn91,bruenn93} & 1987 & No & 2 & 1 (GR)   & 1 & 1 (${\cal O}(v/c)$) \\
\hline
Lawrence Livermore\cite{mayle90,mayle91,wilson93} & 1989 & Yes$^*$ & ``2.25'' & ``1.5'' NS+HR (GR) & 1 & 1 (GR)  \\
\hline
Lawrence Livermore\cite{miller93} & 1992 & Yes$^*$ & 2 & 2 HR (N)  & 2 & 0 (N)  \\
\hline
Los  Alamos\cite{herant94} & 1993  & Yes$^*$ & ``1.75'' & 2 (N)  & ``1.5'' thick/thin & 0 (PN)\\
Arizona\cite{burrows95} & 1994 & Yes$^*$ & ``1.75'' & 2 (N)  & ``1.5'' ray-by-ray & 0 (N) \\
\hline
Florida Atlantic\cite{bruenn94,bruenn95}& 1994  & No & ``2.25'' & ``1.5'' NS (GR) & 1 & 1 (${\cal O}(v/c)$) \\
\hline
Oak Ridge\cite{mezzacappa98a,mezzacappa98b} & 1996 & No & ``2.5'' & 2 (N) & 1 & 1 (${\cal O}(v/c)$)  \\
\hline
Max Planck\cite{rampp00,rampp02,kitaura03,janka04b}  & 2000 & No, Yes$^*$  (ONeMg)& 3 & 1 (N) & 1 & 2 (${\cal O}(v/c)$) \\
Oak Ridge\cite{mezzacappa93b,mezzacappa93c,mezzacappa99,liebendoerfer00,mezzacappa01}  & 2000 & No & 3 & 1 (N) & 1 & 2 (${\cal O}(v/c)$) \\
Arizona\cite{burrows00,thompson03} & 2002 & No & 3 & 1 (N) & 1 & 2 (${\cal O}(v/c)$)  \\
\hline
Oak Ridge\cite{mezzacappa93b,mezzacappa93c,mezzacappa99,liebendoerfer01a,liebendoerfer02,liebendoerfer04} & 2000  & No & 3 & 1 (GR)& 1 & 2 (GR)\\
\hline
Los  Alamos\cite{herant94,fryer02}   & 2002 & Yes$^*$ & ``2.5'' & 3 (N)  & ``2'' thick/thin & 0 (PN) \\
\hline
Max Planck\cite{rampp02,buras03,janka02,janka04a,janka04b}  & 2003 & No, Yes$^*$ (180$^{\mathrm{o}}$)& ``3.75'' & 2 (PN)& ``1.5'' ray-by-ray & 2 (${\cal O}(v/c)$, PN) \\
\hline
\end{tabular}\\[2pt]}
\begin{tabnote}
The ``Yes'' entries in the  ``Explosion'' column are all marked with an asterisk as a reminder that questions about the simulations---described in the main text---have prevented a consensus about the explosion mechanism. 
``Total dimensions'' is the average of ``Fluid space dimensions'' and  ``$\nu$ space dimensions,'' added to ``$\nu$ momentum dimensions.''
The abbreviation ``N'' stands for `Newtonian,' while ``PN''---for `Post-Newtonian'---stands for some attempt at inclusion of general relativistic effects, and ``GR'' denotes full relativity. A space dimensionality in quotes---like ``1.5''---denotes an attempt at modeling higher dimensional effects within the context of a lower dimensional simulation. For the fluid, this is a mixing-length prescription in the neutron star (``NS'') or the heating region (``HR'') behind the stalled shock. For neutrino transport, it indicates one of two approaches: multidimensional diffusion in regions with strong radiation/fluid coupling, matched with a spherically symmetric `light bulb' approximation in weakly coupled regions (``thick/thin''); or the (mostly) independent application of a spherically symmetric formalism/algorithm to separate spatial angle bins (``ray-by-ray'').
\end{tabnote}
\end{sidewaystable}

We pick up the story in 1982, when simulations showing the stalled shock reenergized by neutrino heating on a time scale of hundreds of milliseconds were first performed.\cite{wilson85} This was initially achieved in a simulation with a total of 2 dimensions (spherical symmetry, and energy-dependent neutrino transport). But these simulations required significant rezoning, possibly attended by nontrivial numerical error;\cite{wilson85} and further, with the introduction of full general relativity and a correction in an outer boundary condition,\cite{mayle90} it became clear that these models would not explode without a mock-up of a doubly-diffusive fluid instability in the newly-born neutron star that serves to boost neutrino luminosities\cite{mayle90,mayle91,wilson93,miller93}---a simulation of effective total dimensionality ``2.5'' (see Table 1). That the necessary conditions exist for this particular instability to operate has been disputed;\cite{bruenn95,bruenn96,bruenn04} and though related phenomena may operate,\cite{bruenn04} more recent simulations with energy-dependent neutrino transport and true two-dimensional fluid dynamics indicate that fluid motions are either suppressed by neutrino transport\cite{mezzacappa98a} or have little effect on neutrino luminosities and supernova dynamics.\cite{buras03,janka04a}

Recognizing that the profiles obtained in spherically symmetric simulations implied convective instabilities, and that observations of supernova 1987A also pointed to asphericities, several groups explored fluid motions in two spatial dimensions in the supernova environment in the 1990s. In two spatial dimensions, 
the computational limitations of that era required approximations that
simplified the neutrino
transport. 

One class of simplifications allowed for neutrino transport in ``1.5'' or 2 spatial dimensions, but with
neutrino energy and angle dependence integrated out, 
reducing a five dimensional problem to  ``1.75'' or 2 effective total dimensions (see Table 1).\cite{miller93,herant94,burrows95}
These simulations exhibited explosions, and elucidated an undeniably important physical effect: a negative entropy gradient behind the stalled shock results in convection that increases the efficiency of heating by neutrinos.
However, in the scheme of Table 1, the inability to track the neutrino energy dependence in these simulations could be viewed as a minor step backwards in effective total dimensionality. The energy dependence of neutrino interactions has the important effect of enhancing core deleptonization, which makes explosions more difficult;\cite{bruenn85,bruenn89a,bruenn89b} this raised the question of whether the exploding models of the early- and mid-1990s were too optimistic. 

This concern about the lack of neutrino energy dependence received some support from a simulation in the late 1990s involving 
a different simplification of neutrino transport:
the imposition of energy-dependent
neutrino distributions from spherically symmetric simulations
onto fluid dynamics in two spatial dimensions.\cite{mezzacappa98b}
Unlike the 
simulations
discussed above, these did not 
explode, casting doubt
upon claims that convection-aided
neutrino heating constituted a robust explosion mechanism.

The nagging qualitative difference between spatially multidimensional
simulations with different neutrino transport approximations 
motivated interest in the possible importance of even more complete neutrino
transport: Might the retention of both the energy
{\em and} angle dependence of the neutrino distributions improve the chances of explosion, as preliminary ``snapshot'' studies suggested?\cite{messer98,burrows00}
Of necessity, the first such simulations were performed in
spherical symmetry, which nevertheless represented an advance to a total dimensionality of 3 (see Table 1).
Results from
three different groups are in accord: Spherically symmetric models of iron core collapse 
do not explode, even with solid neutrino transport\cite{rampp00,mezzacappa01,thompson03} and general relativity.\cite{liebendoerfer01a,liebendoerfer04} Recently, however, it has been shown that the more modest oxygen/neon/magnesium cores of the lightest stars to undergo core collapse (8-10 M$_\odot$) may explode in spherical symmetry.\cite{kitaura03,janka04b}

The current state of the art in neutrino transport in supernova simulations determining the long-term fate of the shock has been achieved by a group centered at the Max Planck Institute for Astrophysics in Garching, who deployed their spherically symmetric energy- and angle-dependent neutrino transport capability\cite{rampp02} along separate radial rays, with partial coupling between rays.\cite{janka02} Initial results---from axisymmetric simulations with a restricted angular domain---were negative with regards to explosions (in spite of the salutary effects of convection, and also rotation),\cite{buras03} apparently supporting the results of Ref. \cite{mezzacappa98b}. 
An explosion was seen in one simulation\cite{janka03}
in which
certain terms in the neutrino transport equation corresponding to Doppler shifts
and angular aberration due to fluid motion were dropped; this simulation also yielded a neutron star mass and nucleosynthetic consequences in better agreement with observations than the ``successful'' explosion simulations of the 1990s,\cite{herant94,burrows95} arguably because of more accurate neutrino transport in the case of both observables. 
The continuing lesson
is that getting the details of the neutrino transport right makes a difference.

In addition to accurate neutrino transport, low-mode ($\ell = 1,2$) instabilities that can develop only in simulations allowing the full range of polar angles may make a subtle but decisive difference, as in an explosion recently reported by the Garching group.\cite{janka04a,janka04b} This achievement was presaged by earlier studies of the supernova context, which featured a demonstration of the tendency for convective cells to merge to the lowest order allowed by the spatial domain\cite{herant92} and a newly-recognized spherical accretion shock instability\cite{blondin03} (discovered independently in a different context in Ref. \cite{foglizzo02}). These these global asymmetries may be sufficient to account for observed asphericities that have often been attributed to rotation and/or magnetic fields. 

Surely every `Yes' entry in the explosion column of Table 1 has been hailed in its time as `the answer' (at least by some!), and as a community we cannot help hoping once again that these recent developments mark the turning of a corner; but important work remains to verify if this is the case. Several groups are committed to further efforts. For example, the Terascale Supernova Initiative (TSI, which includes authors of Ref. \cite{blondin03}) comprises efforts aimed at `ray-by-ray' simulations\cite{hix01} like those of the Garching group, as well as full spatially multdimensional neutrino transport, both with energy dependence only\cite{myra04} and with energy {\em and} angle dependence (Sec. \ref{sec:transport}, and Ref. \cite{cardall04}). Delineation of the possible roles of rotation and magnetic fields are also being pursued by TSI. At least one other group is pursuing full spatially multidimensional neutrino transport.\cite{livne04,walder04}

\section{{\em GenASiS:} Philosophy and Basic Features}
\label{sec:genasis}

As discussed in Sec. \ref{sec:challenges}, a core-collapse supernova involves a six-dimensional radiation hydrodynamics problem, making it a major computational challenge. Even three-dimensional pure hydrodynamics problems (with only space dimensions, no momentum space) have only become relatively common in the last few years, with manageable workflows on today's terascale machines ($\sim 10^{12}$ bytes of memory and flop/s). To begin to get a feel for the requirements of {\em radiation} hydrodynamics, consider just a five-dimensional problem, in which axisymmetry in the space dimensions is assumed. For example, supposing the numbers of spatial zones in spherical coordinates $(r,\theta)$ to be $(256,128)$, and the numbers of momentum bins in energy and angle variables $(\epsilon,\vartheta,\varphi)$ to be $(64,32,16)$, of order $10^{10}$ bytes are required just to store one copy of one neutrino distribution function. While this gives rise to a taxing (but not necessarily insurmountable) workflow on terascale machines, it is apparent that the addition of the third spatial dimension---necessary for full exploration of the interacting effects of convection, rotation, and magnetic fields---will require petascale systems. But petascale systems will eventually be available (five to seven years is the current expectation); and given the long development time scales of sophisticated software, we believe it wise to develop our code with the full six-dimensional capability, even if it is only deployed in five dimensions in the near term.

The computational demands of radiation hydrodynamics can be 
ameliorated by ``adapative mesh refinement'' (AMR). The basic idea
of AMR is to employ high resolution only where needed in order to
conserve memory and computational effort. Our current expectation is to allow for refinement only in the space dimensions. This will help with the management of two difficulties: the large dynamic range in length scales associated with the density increase of six orders of magnitude that occurs during core collapse, and adequate resolution of particular features of the flow (the shock, for instance). With Eulerian codes in multiple space dimensions, these tasks require high resolution; and particularly for radiation hydrodynamics, the savings achievable by reducing the number of zones is considerable, since an entire three-dimensional momentum space is carried by each spatial zone. 

Of the two basic types of AMR on structured grids---the block-structured and zone-by-zone varieties---we have chosen the zone-by-zone approach for use with neutrino radiation hydrodynamics.
Block-structured AMR\cite{berger89} involves the deployment of subgrids
of a certain reasonable minimum size (e.g. eight or sixteen zones per side)
at various levels of refinement. A basic solver routine is applied
independently to each subgrid.  Extra spatial zones (referred to
as `guard zones') are required on the edges of each subgrid, which carry
information from neighboring subgrids; these become the boundary conditions
applied by the solver routine. The strategy is designed for explicit
solution algorithms, in which the functions describing time evolution
need only be evaluated at the previous time step. 
However, the
rapid time scales of neutrino interactions with the fluid require an
{\em implicit} solution algorithm, in which the functions describing the
evolution of the neutrino radiation field are evaluated at the {\em current} (that is, new) time step.
Because of this mismatch with the intended purposes of block-structured AMR (implicit vs. explicit evolution), and the fact that popular block-structured AMR community packages did not seem readily amenable to handling momentum space variables in a natural way,
we decided upon another flavor of AMR: the zone-by-zone refinement approach.\cite{khokhlov98} In this
method, individual zones are refined (typically by bisection) and coarsened as 
needed. This provides more flexibility than the block-structured approach; the fine-grained control allows for maximum savings in the number of spatial zones deployed.
A drawback for many users is that a single-grid explicit solver cannot be used ``as is.''
Instead, new solution algorithms must be developed that address the entire hierarchical data structure (Ref. \cite{khokhlov98} and our Sec. \ref{sec:hydro} are examples for explicit hydrodynamics); but we are required to develop such `global solvers' for gravity (Sec. \ref{sec:gravity}) and implicit neutrino transport (Sec. \ref{sec:transport}) anyway. And as a bonus, the need to carry memory-wasting `guard zones' is obviated.

\begin{figure}
\includegraphics[width=3.5in,angle=270]{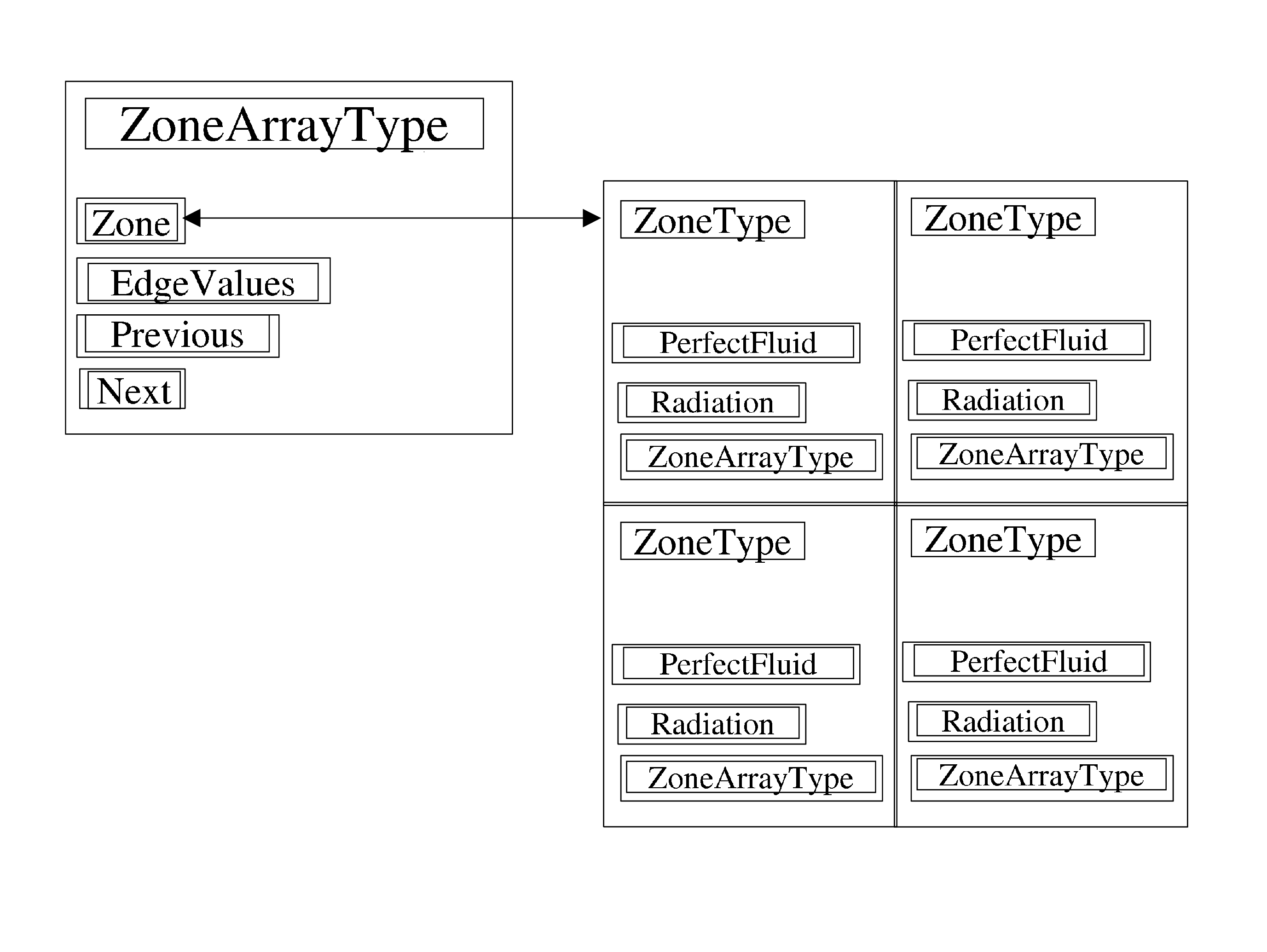}
\caption{Data structures used in an adaptive mesh for radiation 
hydrodynamics.}
\label{zone}
\end{figure}

\begin{figure}
\includegraphics[width=4.7in]{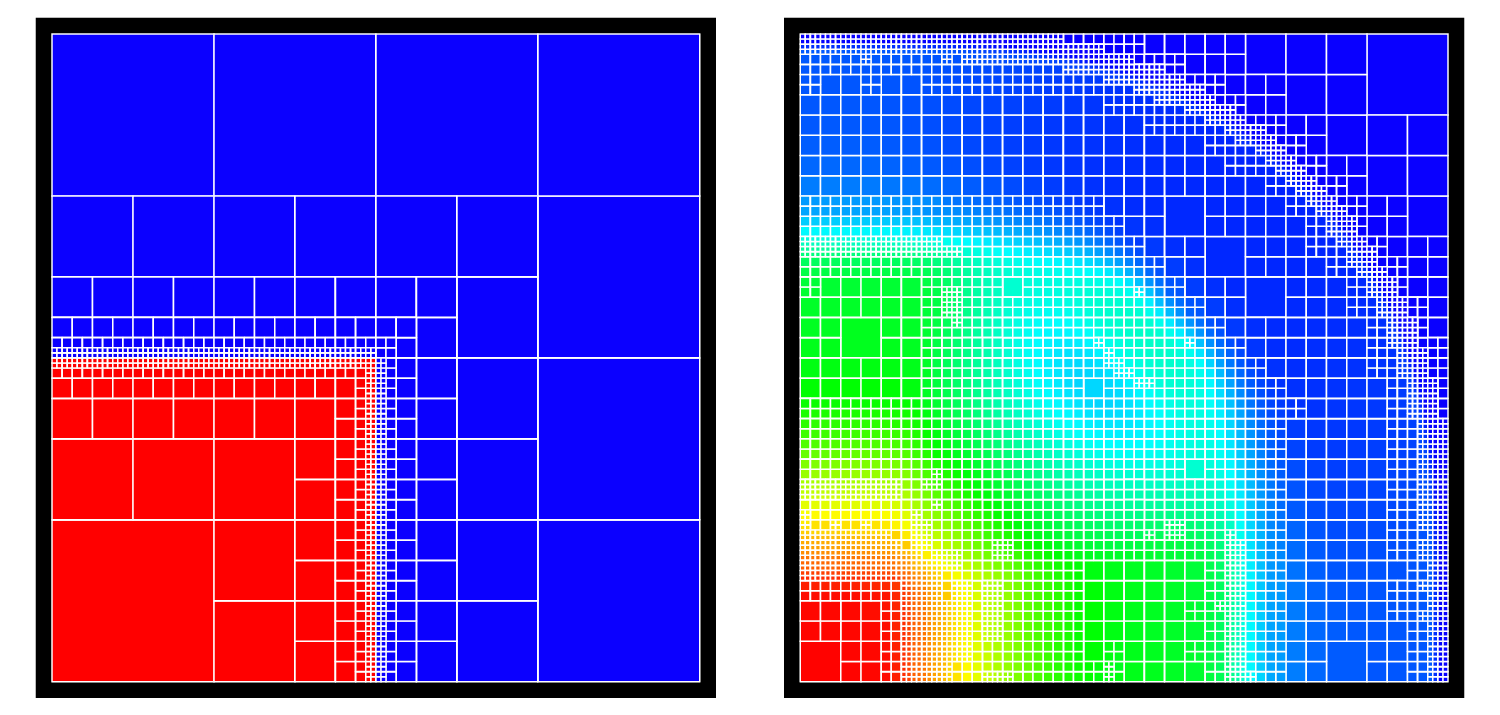}
\caption{Density in a two-dimensional generalization of the shock
tube. Red and blue indicate high and low density respectively. 
Left: Initial state. Right: Evolved state.}
\label{initialFinal2D}
\end{figure}

In implementing the zone-by-zone-refinement approach to the representation of spacetime, we have tried to follow object-oriented design principles to the extent allowed by Fortran 90/95.
Figure \ref{zone}
outlines the basic data structures we use to model the ideal of a 
continuous spacelike slice with a discretized approximation. (The hierarchy of structures, and our operations on them with well-controlled interfaces, are instances of the object-oriented principles of {\em inheritance} and {\em encapsulation}.) A region
of a spacelike slice is represented by an object of {\tt zoneArrayType}. 
Each such object contains an array of objects of {\tt zoneType}, along with  
information about the coordinates of the zones and pointers to neighboring 
zone arrays. Each zone, an object of {\tt zoneType}, contains various forms
of stress-energy, each of which is a separate object. Figure \ref{zone}
shows a perfect fluid and a radiation field; in the code we have an electromagnetic field as well. 
Each zone has a pointer to another object of {\tt zoneArrayType}, whose allocation
constitutes refinement of that zone;
this structure can be extended to arbitrarily deep. A simple two-dimensional pure hydrodynamics test problem computed with our adaptive mesh code is shown in Fig. \ref{initialFinal2D}.

The word ``General'' that goes into the name of our code, {\em GenASiS}---for {\em Gen}eral {\em A}strophysical {\em Si}mulation {\em S}ystem---may give an initial impression of a messianic quest to create an impossibly all-purpose code for solving all conceivable problems in astrophysics and cosmology; but the code's `generality' is, of course, considerably more modest: It 
refers to the use of Fortran 90's facility for function overloading. (This is an instance of the object-oriented principle of {\em polymorphism}.) This allows a generic function name to have several different implementations, providing for extensibility of the physics: Different equations of state, hydrodynamic flux methods, coordinate systems, gravity theories, and so forth can be employed by adding new implementations of generic function names, without having to go back and change basic parts of the code to implement new physics.

\section{Magnetohydrodynamics}
\label{sec:hydro}

We employ a conservative formulation of the equations of magnetohydrodynamics.
The Newtonian case will be described here. Conservation of baryons is described by the equation
\begin{equation}
{\partial n\over\partial t} + \bnabla\cdot(n\bv)=0,\label{baryon}
\end{equation}
where $n$ is the baryon number density, and $\bv$ is the fluid velocity.
The equation 
\begin{equation}
{\partial\over\partial t}(mn\,\bv) + \bnabla\cdot\left[mn\,\bv\bv+\left(p+{B^2\over2}\right)\,\bOne-\bb\bb\right]= -mn\,\bnabla\Phi
\end{equation}
describes conservation of momentum. Here $m$ is the average baryon mass, $p$ is the pressure, $\bb$ is the magnetic field, $\Phi$ is the gravitational potential (discussed further in Sec. \ref{sec:gravity}), and $\bOne$ is the unit tensor.  Units of the magnetic field are chosen such that the vacuum magnetic permeability is unity. One way to express conservation of energy is
\begin{eqnarray}
{\partial\over\partial t}\left[e+{mn\over2}\left(v^2+\Phi\right)+{B^2\over2}\right] +&&\nonumber\\ \bnabla\cdot\left[(e+p+B^2)\bv+{mn\over2}\left(v^2+\Phi\right)\bv-\bb(\bv\cdot\bb)\right]&=&-{m n\over 2}\left(\bnabla\cdot\bPsi+\bv\cdot\bnabla\Phi\right)-\nonumber\\
& &n\left({\partial m\over\partial t}-\bv\cdot\bnabla m\right),\label{energy}
\end{eqnarray}
where $e$ is the internal energy density, and $\bPsi$ is a kind of ``gravitational vector potential'' (also discussed in Sec. \ref{sec:gravity}). We have opted to employ the baryon number density, instead of the mass density, in explicit deference to the fact that mass is not conserved in the presence of nuclear reactions. The energy input from nuclear reactions has then resulted naturally in the derivation of Eq. (\ref{energy}), appearing in the last two terms of the right-hand side. We hope to add nuclear reaction networks to our code in the future.

This formulation is called ``conservative'' because volume integrals of the divergences in Eqs. (\ref{baryon})-(\ref{energy}) are related to surface integrals through the divergence theorem:
\begin{equation}
\int_V dV\,(\bnabla\cdot \bF) = \oint_{\partial V} \bF\cdot \dA.
\end{equation}
The physical meaning of a conservative equation is that (modulo source terms) the time rate of change of a conserved quantity in a volume is equal to a flux $\bF$ through the volume's enclosing surface. This meaning is built into the finite-difference representation of Eqs. (\ref{baryon})-(\ref{energy}); divergences are represented in discrete correspondence to their mathematical definition, using zone volumes $V$ and face areas $A$:
\begin{equation}
\bnabla\cdot \bF\rightarrow{1\over V_{\lra}}\sum_q\left[\left(A_q\,F^q\right)_{\qra} - \left(A_q\,F^q\right)_{\qla}\right].\label{discreteDivergence}
\end{equation}
Here $q$ runs over the three space dimensions. A double-headed arrow ($\lra$) indicates evaluation at a zone center. Left-arrows ($\qla$) and right-arrows ($\qra$) denote evaluation at zone inner and outer faces respectively, in the $q$ direction; dimensions other than $q$ are evaluated at the zone centers.
In this way the divergence theorem is replicated in every zone, ensuring global conservation to machine precision. Use of generalized zone volumes and areas in Eq. (\ref{discreteDivergence}) enables the use of curvilinear coordinates (``ficticious forces'' arising from curvilinear coordinates must also be included in the momentum equation).

The evolution of the magnetic field is described by Faraday's law:  
\begin{equation}
{\partial \bb\over\partial t} = -\bnabla\times\be, \label{faraday}
\end{equation}
supplemented by the constraint
\begin{equation}
\bnabla\cdot\bb = 0. \label{divB}
\end{equation}
In Eq. (\ref{faraday}), we take the electric field to be $\be = -\bv\times\bb$, in accordance with the usual astrophysical assumption of a perfectly conducting medium.

While Eq. (\ref{faraday}) for the evolution of the magnetic field does not require conservation of the magnetic field, Eq. (\ref{divB}) requires the magnetic field to be divergence-free at all times.  In the presence of discontinuous flow numerical solutions to Eqs. (\ref{baryon})-(\ref{faraday}) can produce severe unphysical artifacts if this requirement is not met,\cite{brackbill80}
but it can be automatically enforced by the method of constrained transport.\cite{evans88} Integrating Eq. (\ref{faraday}) over a zone's enclosing surface, the left-hand side becomes the volume integral of $\bnabla\cdot\bb$, via the divergence theorem. The right-hand side is a sum over area integrals over each zone face. With Stokes' theorem,  
\begin{equation}
\int_A (\bnabla\times \be)\cdot \dA = \oint_{\partial A} \be\cdot\dl, \label{stokes}
\end{equation}
each of these surface integrals becomes a line integral around the zone face boundary. Summed over all faces, two line integrals in opposite directions cancel on every zone edge, enforcing $\bnabla\cdot\bb = 0$ in the zone as desired.
The method of constrained transport, then, is to evaluate $\bnabla\times\be$ on zone edges in discrete correspondence to the mathematical definition of the curl, using zone face areas $A$ and edge lengths $L$:
\begin{equation}
(\bnabla\times\be)_q \rightarrow{1\over A_q}\sum_{r\ne q}\left[\left(L_s\,E^s\right)_{\rra} - \left(L_s\,E^s\right)_{\rla}\right].\label{discreteCurl}
\end{equation}
Here $r$ runs over the two space dimensions orthogonal to a particular direction $q$, and $s$ indicates the direction perpendicular to both $q$ and $r$. Left-arrows ($\rla$) and right-arrows ($\rra$) denote evaluation along face inner and outer edges respectively, in the $r$ direction.
This ensures divergence-free evolution of the discrete representation of the area-averaged magnetic field, with components located on the appropriate zone faces, for all times to machine precision, provided the initial magnetic field satisfies Eq. (\ref{divB}).  

Accurate computation of fluxes at zone faces and electric fields at zone edges is a key feature. So-called ``central schemes'' have been noted recently by astrophysicists for their ability to capture shocks with an accuracy comparable to Riemann solvers, but with much greater simplicity.\cite{lucas04} In particular, we employ so-called ``HLL'' versions of these schemes for both fluid conservation laws\cite{delZanna02} and the magnetic induction equation.\cite{delZanna03} We achieve second order in space by linear interpolation within zones (as usual, a slope limiter---deployed where necessary in order maintain discontinuities---reduces the treatment to first order).

While we have taken Khokhlov's zone-by-zone refinement approach\cite{khokhlov98} as our basic paradigm, we have made a novel extension to evolve the magnetic field, and use a different time-stepping scheme. As in Khokhlov's work, fluxes are computed at each zone interface only once, with the results used to update zones on both sides of the interface. At coarse/fine interfaces, fluxes are computed only on the faces of the refined zones. We have developed a similar approach for the induction equation: the electromotive forces on the zone edges are computed only once, and are used to update all zones sharing that edge. We use a second-order Runge-Kutta time stepping algorithm, made possible by the semi-discrete formulation of the central scheme. In doing so we evolve all levels of the mesh synchronously, unlike Khokhlov's approach of evolving refined levels with greater frequency. In addition to making it possible to take advantage of the semi-discrete formulation for time evolution, problems we encountered in self-gravitating systems with `asynchronous' evolution a la Khokhlov were avoided.

Parallelization is achieved by giving each processor its share of spatial zones. Partitioning is accomplished by walking through all levels of the mesh in a recursive manner similar to a Morton space-filling curve; the result is a mapping of the multidimensional mesh to a one dimensional ``string'' of zones, which, when cut into pieces of uniform length, leaves each processor with roughly the same number of zones at each level of refinement.

An equation of state determines $p$, the quantity in Eqs. (\ref{baryon})-(\ref{energy}) whose determination has not yet been mentioned. So far, the code has two overloaded options. One is the familiar polytropic equation of state:
\begin{eqnarray}
p&=&\kappa\, n^\Gamma, \\
e&=&(\Gamma-1)^{-1} p,
\end{eqnarray}
where $\Gamma$ is a specified parameter, and $\kappa$ is updated in response to changes in $e$ determined from Eq. (\ref{energy}). We have also implemented a ``realistic'' equation of state\cite{lattimer91} suitable for problems involving nuclear matter. This equation of state takes as input the temperature, baryon number density, and electron fraction $Y_e$, defined by
\begin{equation}
Y_e = {n_{e^-} - n_{e^+}\over n},
\end{equation}
where $n_{e^-}$ and $n_{e^+}$ are the number densities of electrons and positrons respectively. When using this ``realistic'' equation of state, an advection equation for $Y_e$ must be added to the above list of conservation laws.

\begin{figure}
\begin{center}
$\begin{array}{@{\hspace{-0.25in}}c@{\hspace{-0.75in}}c}
\includegraphics[width=3in]{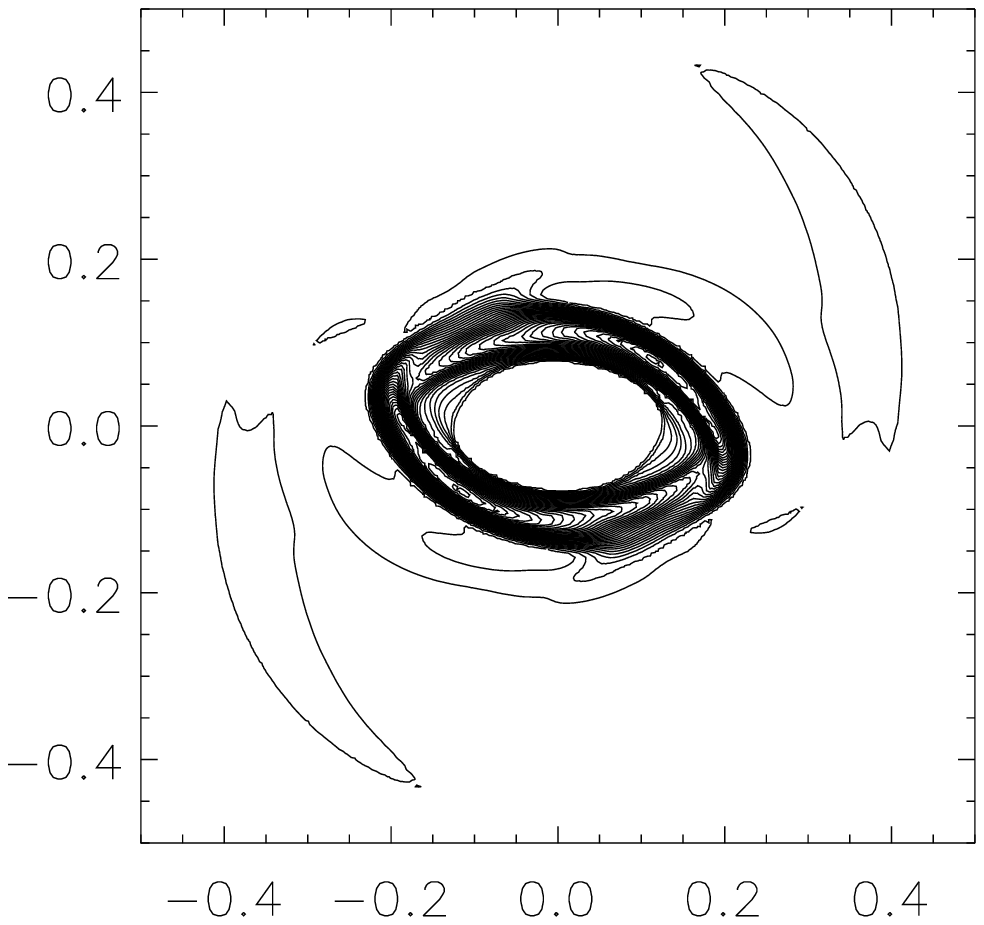} &
\includegraphics[width=3in]{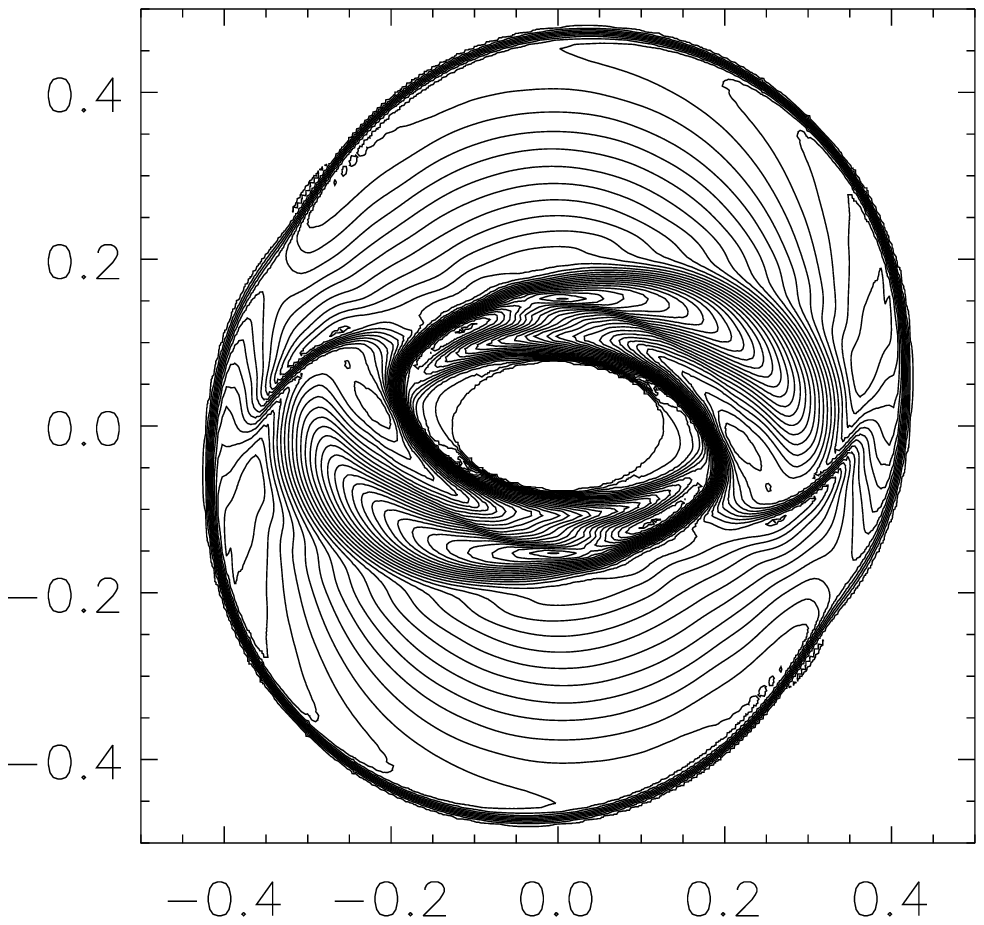} 
\end{array}$
\end{center}
\caption{Density (left panel) and thermal pressure (right panel) at $t=0.295$ for the second rotor problem given in Ref. \protect\cite{toth00}.  40 countours were used to produce the plots, with $0.512 \le mn \le 9.622$ and $0.010 \le p \le 0.776$.  A $200\times 200$ grid was used to produce the results.}
\label{rotorContours}
\end{figure}

We have been working with a number of hydrodynanic and magnetohydrodynamic test problems, one of which is shown here. 
The rotor problem---which consists of a rapidly rotating dense fluid, initially cylindrical, threaded by an initially uniform magnetic field---was devised to test the onset and propagation of strong torsional Alfv\'en waves into the ambient fluid.\cite{balsara99}
We have computed a version of the rotor problem with initial data identical to a so-called 'second rotor problem,'\cite{toth00} and display the results in Fig. \ref{rotorContours}.  

\section{Newtonian Gravity}
\label{sec:gravity}

The Poisson equation for the Newtonian gravitational potential $\Phi$ is 
\begin{equation}
\nabla^2\Phi = 4\pi G\, m n,\label{poisson}
\end{equation}
where $G$ is the gravitational constant. Our finite-differenced approach to Eq. (\ref{poisson}) is based on the fact that the Laplacian is a divergence of a gradient:
\begin{equation}
\bnabla\cdot\bnabla\Phi = 4\pi G\, m n.\label{poisson2}
\end{equation}
We discretize this equation in a manner similar to Eq. (\ref{discreteDivergence}). There results a linear system for the values of $\Phi$ at the center of every leaf zone. In the matrix representation of this linear system, each row corresponds to the discrete version of Eq. (\ref{poisson2}) centered on a given ``leaf zone'' (those not having refined children). On a single-level grid, the resulting matrix would have three, five, and seven bands in one, two, and three dimensions respectively. With adaptive mesh refinement, the matrix structure becomes more diffuse, because several refined zones may contribute to the discrete representation of $\bnabla\Phi$ at a zone face featuring a coarse/fine interface. Each processor fills in the portion of the matrix corresponding to its share of zones, and we rely on the PETSc library ({\tt http://www-unix.mcs.anl.gov/petsc/petsc-2/}) to perform the distributed sparse matrix inversion. 

We now discuss the contribution of gravity to the fluid energy evolution in Eq. (\ref{energy}). Without the gravitational potential energy included in the ``total energy'' in the time derivative, Eq. (\ref{energy}) appears in the more familiar form
\begin{eqnarray}
{\partial\over\partial t}\left[e+{1\over2}mn\,v^2\right] +\bnabla\cdot\left[(e+p+{1\over2}mn\,v^2)\bv\right]&=&-m n \,\bv\cdot\bnabla\Phi-\nonumber\\
& &n\left({\partial m\over\partial t}-\bv\cdot\bnabla m\right).\label{energy2}
\end{eqnarray}
Including the graviational potential energy density $(m n/2)\Phi$ in the time derivative, and making use of baryon conservation, we have
\begin{eqnarray}
{\partial\over\partial t}\left[e+{mn\over2}\left(v^2+\Phi\right)\right] +&&\nonumber\\ \bnabla\cdot\left[(e+p)\bv+{mn\over2}\left(v^2+\Phi\right)\bv\right]&=&{m n\over 2}\left({\partial\Phi\over\partial t}-\bv\cdot\bnabla\Phi\right)-\nonumber\\
& &n\left({\partial m\over\partial t}-\bv\cdot\bnabla m\right).\label{energy3}
\end{eqnarray}
Using the formal solution for $\Phi$,
\begin{equation}
\Phi(\bx,t)= - G\int {m(\bx',t) n(\bx',t)\,d^3 x'\over |\bx-\bx'|},
\end{equation}
together with baryon conservation (and neglecting time derivatives of $m$), one can show that
\begin{equation}
{\partial\Phi\over\partial t} = -\bnabla\cdot\bPsi, 
\end{equation}
where $\bPsi$ satisfies a vector Poisson equation:
\begin{equation}
\nabla^2\bPsi = 4\pi G\, m n\, \bv.
\end{equation}
This may be solved in a manner similar to that used to solve for $\Phi$ (though the vector Poisson equation contains some additional terms in curvilinear coordinates).
Conservation of total energy, including gravitational, is not local: the gravitational source terms (the first two terms on the right-hand side of Eq. (\ref{energy})) vanish only upon integration over all space.

\section{Neutrino Radiation Transport}
\label{sec:transport}

Here we briefly describe our approach to the greatest computational challenge in supernova simulations: neutrino radiation transport. Neutrino distributions must be tracked in order to compute the transfer of lepton number and energy between the neutrinos and the fluid. There are three major challenges. One challenge is constructing a discretization that allows both energy and lepton number to be conserved to high precision. The two other challenges are associated with the limits of computational resources: the solution of a very large nonlinear system of equations, and neutrino interaction kernels of high dimensionality.

Energy conservation is an obvious measure of quality control, and care with the transport formalism and differencing can help achieve it. The importance of energy conservation is brought into focus by this question: How should we interpret the prediction of a $\sim 10^{51}$ erg explosion in a model where the total energy varies during the course of the simulation by $\sim 10^{51}$ erg or more? To achieve the required precision (say, global energy changes of less than $\sim 10^{50}$ erg), conservative formulations are a useful starting point, and relativistic treatments avoid quantitatively non-negligible conflicts at $O(v^2/c^2)$ between the number and energy transport equations.\cite{liebendoerfer01b,liebendoerfer04,cardall03,cardall05b} Finite-differencing that simultaneously satisfies energy and lepton number conservation has been implemented in spherical symmetry,\cite{liebendoerfer04} and should be pursued in multiple spatial dimensions as well.

Our algorithm for solving the large nonlinear system has been described elsewhere.\cite{dazevedo05,cardall04} The large system of equations requiring inversion (as opposed to explicit updates) results from the disparity between hydrodynamic and particle interaction time scales, which motivates implicit time evolution. The nonlinear solve is achieved with the Newton-Raphson method. A fixed-point method employing a preconditioner that splits the space and momentum space couplings is used for the linear solve required within each Newton-Raphson iteration.\cite{dazevedo05} An advantage of this linear solver method is that the dense blocks representing couplings in momentum space---which cannot all be stored at once---need only be constructed a few at a time, used in all steps required in a given fixed-point iteration, and discarded. In contrast, other linear solver algorithms seem to require dense blocks to be discarded and rebuilt multiple times in each iteration. We have sucessfully tested this solver on a two-dimensional problem in spherical coordinates with a static background and a simple emission/absorption interaction.

Because neutrino interactions are expensive to compute on-the-fly, we have implemented interpolation tables. Neutrino interactions depend on the neutrino momentum components and the state of the fluid with which the neutrinos interact. A grid of neutrino energies and angles is fixed, but the fluid density $n$, temperature $T$, and electron fraction $Y_e$ vary throughout the simulation. Hence we employ tables that may be interpolated in $n$, $T$, and $Y_e$. Particularly for neutrino scattering and pair interactions---which depend on neutrino states before and after the collision---the interaction kernels are of high dimensionality, requiring a globally distributed table. On each processor a local table is constructed, which contains a copy of each $n$, $T$, and $Y_e$ vertex required by the zones for which that processor is responsible. As $n$, $T$, and $Y_e$ in a processor's zones evolve, the relevant vertices are pulled from the global table as needed. For each zone we construct a ``cube'' of pointers to the eight vertices surrounding the zone's values of $n$, $T$, and $Y_e$. This cube is then used for the necessary interpolations.

\section{Outlook}

We have made a promising start on {\em GenASiS}, a new code being developed to study the explosion mechanism of core-collapse supernovae. Our plan is to include all the relevant physics---including magnetohydrodynamics, gravity, and energy- and angle-dependent neutrino transport---in a code with adaptive mesh refinement in two and three spatial dimensions. Parallelization and implementation on the adaptive mesh are not yet complete, and the physics components have not yet been fully integrated; but steady progress and the successful completion of test problems give us confidence that we are well on our way towards a tool that will provide important insights into the supernova explosion mechanism.

\section*{Acknowledgments}
We gratefully acknowledge S.~W. Bruenn's contribution of subroutines for the computation of neutrino interaction kernels. E.~J. Lentz collaborates on the development of a comprehensive neutrino radiation transport discretization scheme, still too immature to report here. We  thank R.~D. Budiardja and M.~W. Guidry for discussions on the Poisson solver, and R.~D. Budiardja for helping with that solver's interface to the PETSc library. 
This work was supported 
by  Scientific Discovery Through
Advanced Computing (SciDAC), a program of the Office of Science of the U.S. Department of Energy (DoE); and by Oak Ridge National Laboratory, managed by UT-Battelle, LLC, for the DoE under contract DE-AC05-00OR22725.

\def\aap{{\em Astron. Astrophys.}}
\def\aasma{{\em Am. Astron. Soc. Meet. Abs.}}
\def\apj{{\em Astrophys. J.}}
\def\apjl{{\em Astrophys. J. Lett.}}
\def\apjs{{\em Astrophys. J. Suppl. Ser.}}
\def\araa{{\em Annu. Rev. Astron. Astrophys.}}
\def\baas{{\em Bull. Am. Astron. Soc.}}
\def\jcam{{\em J. Comput. Appl. Math.}}
\def\jcp{{\em J. Comput. Phys.}}
\def\nat{{\em Nature}}
\def\npa{{\em Nucl. Phys. A}}
\def\pr{{\em Phys. Rep.}}
\def\prd{{\em Phys. Rev. D}}
\def\prl{{\em Phys. Rev. Lett.}}
\def\prv{{\em Phys. Rev.}}
\def\rmp{{\em Rev. Mod. Phys.}}


\end{document}